\documentclass[
showpacs,
amsmath,amssymb,
aps,
twocolumn,
prb,
floatfix,
nofootinbib
]{revtex4-2}
\usepackage{xcolor}

\usepackage{graphicx}
\usepackage{amssymb}
\usepackage{dcolumn}

\begin{document}
	
\title{Buzdin, Shapiro and Chimera Steps in  $\varphi_0$ Josephson Junctions
}
\author{Yu. M. Shukrinov $^{1,2,3}$, E. Kovalenko$^{4}$, J. Teki\' c$^{5}$,  K. Kulikov$^{1,2}$, and M. Nashaat$^{1,6}$}
	
\affiliation{$^1$\mbox{BLTP, JINR, Dubna, Moscow region, 141980, Russia}\\
		$^2$ \mbox{Dubna State University, Dubna, Russia}\\	
		$^3$  \mbox{Moscow Institute of Physics and Technology, Dolgoprudny 141700, Russia}\\
		$^4$  \mbox{Center for the Development of Digital Technologies, Krasnogorsk, Russia} \\
		\vbox{$^5$ "Vinča" Institute of Nuclear Sciences, Laboratory for Theoretical \\ and Condensed Matter Physics - 020, University of Belgrade, PO Box 522, 11001 Belgrade, Serbia}\\
		$^6$ \mbox{Department of Physics, Faculty of Science, Cairo University, 12613, Giza, Egypt}
}
		
\date{\today}
	
\begin{abstract}
The unique resonance and locking phenomena in the superconductor-ferromagnet-superconductor $\varphi_0$ Josephson junction under external electromagnetic radiation are demonstrated when not just the electric but also the magnetic component of external radiation is taken into account. Due to the coupling of superconductivity and magnetism in this system, the magnetic moment precession of the ferromagnetic layer caused by the magnetic component of external radiation can lock the Josephson oscillations, which results in the appearance of a particular type of steps in the current-voltage characteristics, completely different from the well-known Shapiro steps. We call these steps the Buzdin steps in the case when the system is driven only by the magnetic component and the Chimera steps in the case when both magnetic and electric components are present. Unlike the Shapiro steps where the magnetization remains constant along the step, here it changes though the system is locked.
The spin-orbit coupling substantially contributes to the amplitude, i.e., the size of these steps. Dramatic changes in their amplitudes are also observed at frequencies near the ferromagnetic resonance. Combinations of the Josephson and Kittel ferromagnetic resonances together with different types of locking pronounced in dynamics and current-voltage characteristics make the physics of this system very interesting and open up a series of new applications.
\end{abstract}
	
\maketitle
The possibility to combine superconductivity and magnetism in hybrid Josephson structures holds promise to increase the technological applications of superconductors and superconducting nanostructures in the recent rapid development of spintronics and superconducting logic devices. One particular structure that demonstrates  transport properties with disrupting scientific and technological potential is the superconductor-ferromagnet-superconductor (SFS) $\varphi_0$ Josephson junction (JJ)~\cite{buz08,kon09}.
It belongs to a special class of anomalous JJs with a noncentrosymmetric ferromagnetic layer and broken time-reversal symmetry, which results in a particular current-phase relation $I=I_c\sin(\varphi -\varphi _0)$ with the phase shift proportional to the magnetic moment \cite{shuk22ufn}.
Experimental observations of this anomalous phase shift in different systems~\cite{szom16,apr19,may19} open up several new opportunities for superconducting spintronics~\cite{lin15}. The presence of {\it bidirectional} coupling between the magnetic moment of the barrier and the superconducting phase difference allows superconductivity to control magnetism and vice versa, to influence Josephson current via magnetic moment, which could lead to a series of new applications~\cite{buz08,kon09,lin15,shuk22ufn,mel22ufn,bobkova-rev}.
	
\begin{figure}[tbh]
		\centering
		\includegraphics[height=30mm]{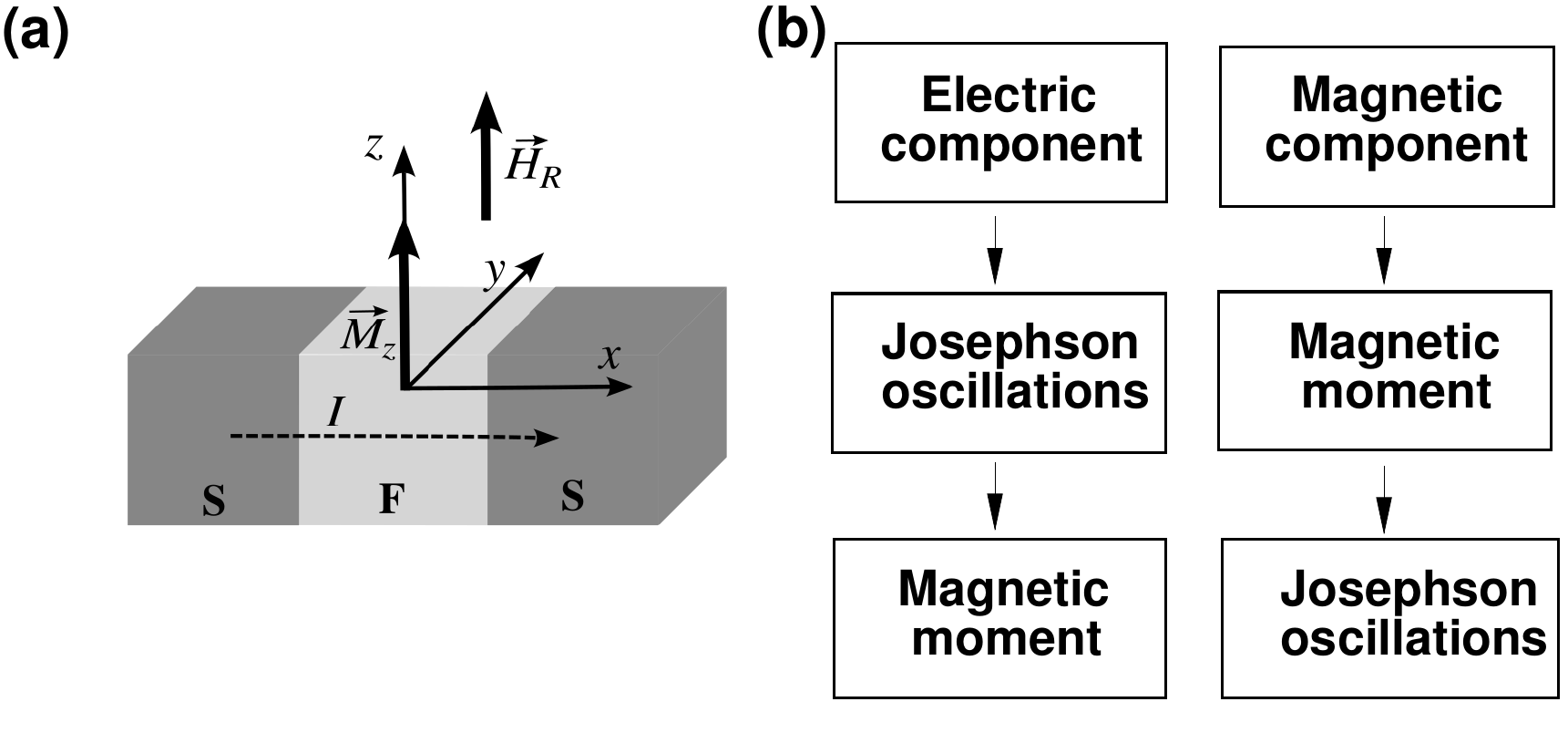}
		\caption{(a) Geometry of the system; (b) Demonstration of two locking mechanisms in the $\varphi_0$ Josephson junction under an external electromagnetic field (see the text).}
		\label{Fig1}
\end{figure}

In the studies of ordinary superconductor-insulator-superconductor (SIS-type) Josephson junctions driven by external radiation, the influence of the magnetic component of radiation is usually neglected, and the description of the effect is reduced to adding the term $A sin (\omega t)$ to the bias current. As was mentioned in Ref.\cite{kon09}, in the $\varphi_0$ JJ  the microwave magnetic field generates an additional magnetic precession with the microwave frequency which might lead to a series of unusual effects. However, so far these predictions have not been verified and detailed studies of the interaction of electromagnetic radiation with the $\varphi_0$-junction taking into account the magnetic component have not been carried out. Here we eliminate this shortcoming and include the direct interaction of the magnetic component of the microwave magnetic field  with the magnetic moment of the ferromagnetic layer in our study of radiation effects in the $\varphi_0$ junction. The considered geometry is demonstrated in Fig.\ref{Fig1}(a).
	
In this Letter, we demonstrate the effects of microwave radiation on the dynamics and the $IV$-characteristics of the SFS $\varphi_0$ Josephson junction taking into account both the magnetic and electric components of radiation.
This leads to two different mechanisms of locking the Josephson oscillations and the ferromagnetic moment precessions (see  Fig.\ref{Fig1}(b)).  These mechanisms of indirect locking are investigated at different parameters of the $\varphi_0$ junction and microwave field.
Additionally to the Shapiro steps caused by the electric component of radiation, we observe a unique step created by the periodic field of the magnetic component.  By locking the magnetic moment precession, one can also lock the Josephson oscillations due to their coupling with the ferromagnetic moment. We call this step the Buzdin (BS) step since it was first predicted in Ref.\cite{kon09} to stress its different origin and properties from the Shapiro step (SS). When both radiation components are taken into account, we observe another type of step named the composite or Chimera step due to its creation by two different mechanisms. Below we will call this step the C-step (CS). We show that CS is not a trivial sum of BS and SS; however, it has its own specific features reflecting its origin from two different locking mechanisms. While the electric component interacts with the superconducting current, the magnetic component, by driving the ferromagnetic moment, interacts also with the superconducting subsystem due to the coupling of Josephson oscillations and ferromagnetic moment precession. 
We also show that when either Josephson or radiation frequencies approach the ferromagnetic one the interplay of the Josephson and Kittel ferromagnetic resonances appears.
	
The model is described by the system of equations obtained from the Landau-Lifshitz-Gilbert (LLG) equation, the Josephson relation for the phase difference and voltage, and the equation for the biased current of the resistively and capacitively shunted junction (RCSJ) model:
\begin{eqnarray}
		\frac{d\mathbf{M}}{dt}&=&-\gamma \mathbf{M}\times H_{eff}+\frac{\alpha }{M_0}\left( \mathbf{M}\times \frac{d\mathbf{M}}{dt}\right),\\  \nonumber
		\dot{\varphi}&=&V(t), \\
		\dot{V}&=& \left[ I+A\sin(\omega_Rt)-V(t)+r\dot{m}_{y}-\sin (\varphi -rm_{y})\right]/\beta_c, \nonumber
		\label{LLG}
\end{eqnarray}
with the effective magnetic  field $H_{eff}$:
\begin{eqnarray}
		\textbf{H}_{eff}&=
		&\frac{K}{M_0}Gr\sin \left( \varphi - r\frac{M_y}{M_0}\right)\hat{\textbf{y}} + \\  \nonumber
		& &+\left( \frac{K}{M_0}\frac{M_z}{M_0} + H_R\sin(\Omega_Rt)\right)\hat{\textbf{z}},
		\label{Heff}
\end{eqnarray}
where $\beta_c$ is McCumber parameter, ${G=E_J/(K\mathcal{V})}$ represents the ratio of the Josephson to magnetic anisotropy energy, $r$ is the Rashba spin-orbit coupling (SOC).
The second term inside the sine function represents the phase shift $\varphi _0=rM_y/M_0$.
The gradient of the spin-orbit potential is along the easy axis, which is taken to be along $z$. In the LLG equation $\gamma $ is the gyromagnetic ratio,  $\alpha $ is the Gilbert damping, $M_0=|\mathbf{M}|$.
	
We use the dimensionless variables $m_{i}=\frac{M_{i}}{M_0}, (i\equiv x,y,z), \ \ t\rightarrow t\omega _c, \ \ \omega _R=\frac{\Omega _R}{\omega _c}$, where $\omega _c=2eI_cR/\hbar $ is a characteristic frequency of the junction, the ferromagnetic resonance frequency $\Omega _F = K\gamma /M_0$ and the amplitudes of the electric $\mathbf{E_R}=(E_R\sin (\Omega _Rt),0,0)$, and magnetic $\mathbf{H_R}=(0,0,H_R\sin(\Omega_Rt)$ components  are then  normalized to $\omega _c$, so that  $\omega _F= \frac{\Omega _F}{\omega _c}, \ \  h_R=\frac{\gamma }{\omega _c}H_R$.
	
First, we examine the effects of the magnetic component of external radiation (MCR) on the magnetization dynamics and the $IV$-characteristics in the ferromagnetic resonance region when ferromagnetic resonance frequency is close to the Josephson one, i.e., when $\omega _F\approx \omega _J$.
Unlike the electric component, the magnetic one can interact directly with the magnetic moment of the ferromagnetic barrier, which further leads to the appearance of the Buzdin step in the $IV$-characteristics.
To get an insight into the origin of BS, we switch off the term $A sin (\omega_R t)$, describing the electric component of radiation, and investigate the effect of $h_R$ only, i.e., concentrate on the features, produced by MCR.
		
\begin{figure}[tbh]
		\centering
		\includegraphics[width=0.8\linewidth]{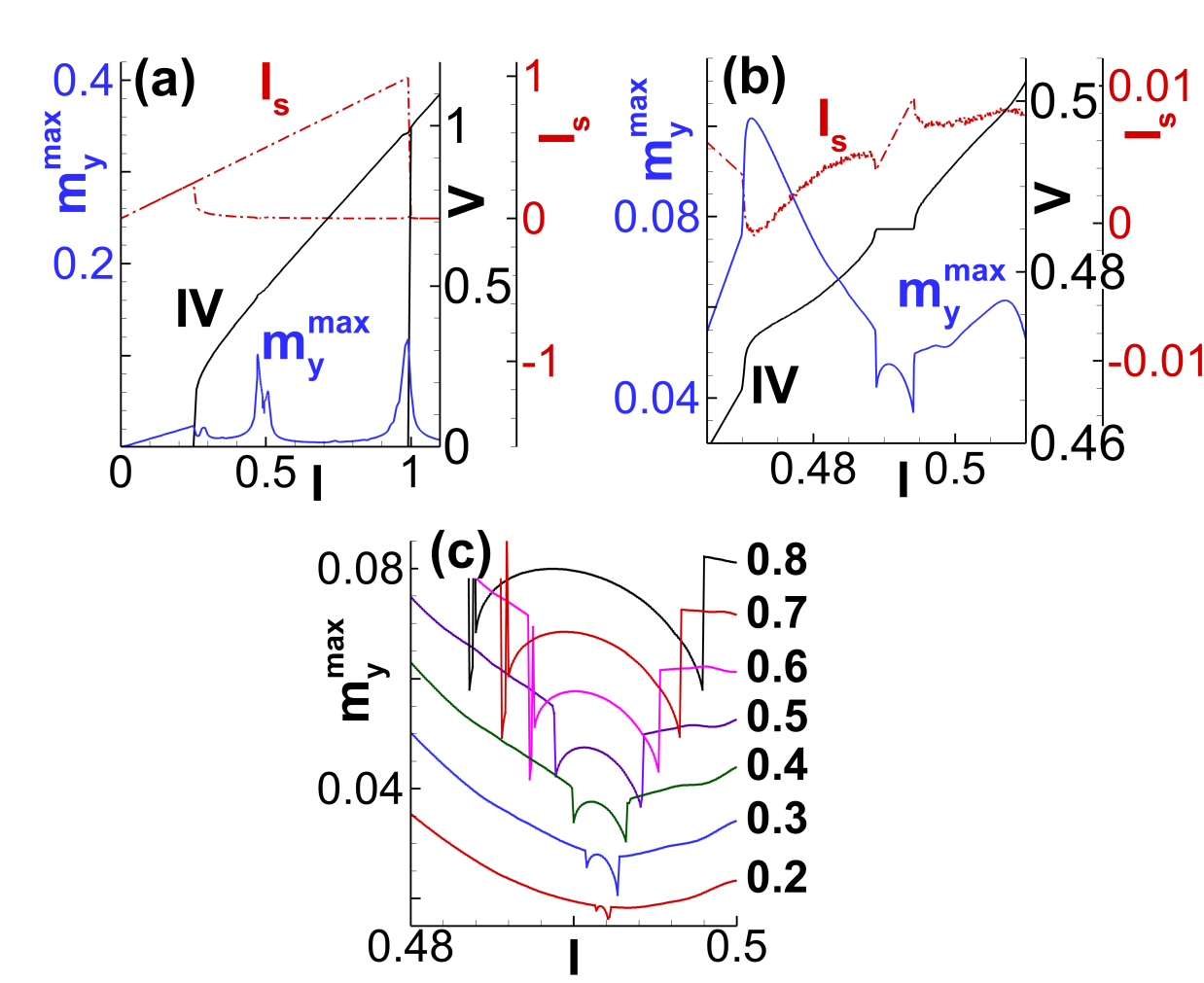}
		\caption{The average voltage $V$, the maximum value of the $m_y$ magnetic component $m_y^{max}$ and the superconducting current $I_s$ as a function of decreasing biased current $I$ for $A=0, h_R=1, r=0.5, G=0.01, \alpha =0.01, \omega _R=0.485, \omega _F=0.5$;  (b) Magnified view of (a) showing the Buzdin step. (c) $m_y^{max}$ at different values of $r$.
		}
		\label{Fig2}
\end{figure}

The average voltage $V$, the maximum value of the magnetic moment, $m_y^{max}$, and the superconducting current $I_s$ as functions of decreasing biased current $I$ are presented in Fig. \ref{Fig2}(a). It shows the ferromagnetic resonance and its manifestation in these characteristics.
The magnified view of Fig. \ref{Fig2}(a) in the FMR region with the Buzdin step is presented in Fig.\ref{Fig2}(b).
Its appearance is a result of the locking of Josephson oscillations by the magnetic component of external radiation. Namely, the MCR creates the periodic precession of the magnetic moment, which then, through coupling with Josephson oscillations, also locks the Josephson oscillations.
So, due to their bidirectional coupling, both the Josephson oscillations and the magnetic precession are locked with MCR. The specific manifestation of this locking is also seen in the  $m_y^{max}(I)$ and $I_s(I)$ in the current interval corresponding to the Buzidn step, where the magnified view reveals a "bubble-like" feature due to changing of $m_y^{max}$ along the step. So we observe the locking of the magnetic moment precession with a changed amplitude but fixed frequency, which is demonstrated below.
	
An interesting question is related to the variation of the parameters characterizing the interaction of Josephson oscillations with the magnetic moment, in particular, the spin-orbit coupling $r$.
This parameter plays a key role in the appearance of the Buzdin steps since it is the coupling between the magnetic moment and Josephson oscillation through which the locking of the magnetic moment transfers to the superconducting subsystem.
Of course, if $r=0$ and there is no coupling, and the Buzdin steps do not exist.
In Fig. \ref{Fig2}(c), the variation  $m_y^{max}(I)$ at different parameters of spin-orbit coupling is presented, where the focus was on the changes of the bubble structure under $r$.
We know from Fig.\ref{Fig2}(b) that the width of the bubble exactly corresponds to the width of the Buzdin step, and as the $r$ dependence shows the bubble, i.e., the Buzdin step increases with the increase of the spin-orbit coupling.
	
\begin{figure}[tbh]
		\centering
		\includegraphics[width=\linewidth]{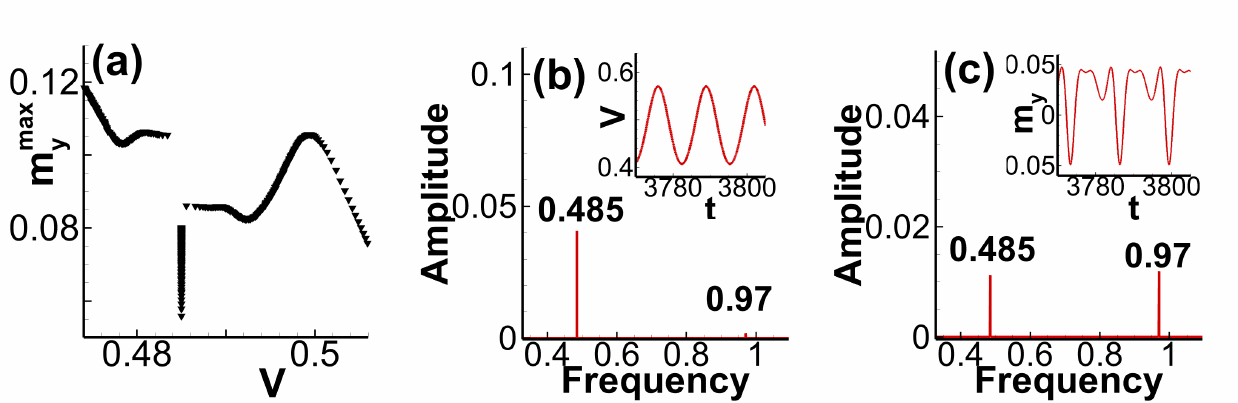}
		\caption{ (a) $m_y^{max}$ as a function of $V$;
			(b) and (c) The time dependence of $V$ and $m_y$, and the corresponding FFT analysis in the center of the bubble, respectively.
			Other parameters are as in Fig. \ref{Fig2}.
		}
		\label{Fig3}
\end{figure}

The proof of the locking by MCR can be seen in Fig. \ref{Fig3} (a), where the voltage dependence of $m_y^{max}$ is presented.
As we can see, there is a sharp minimum at the same value of voltage (i.e., frequency) corresponding to the Buzdin step.
The time dependence of $V$ and $m_y$ and the corresponding Fast Fourier Transform (FFT) analysis
in Fig. \ref{Fig3}(b) and (c), respectively, further confirm the locking by MCR.

Let us now discuss the effects of both electric and magnetic components.  In this case, the electromagnetic irradiation of $\varphi_0$ JJ leads to the realization of two different mechanisms of locking. The electric component of radiation locks the Josephson oscillation, and it, due to the coupling, locks the  precession of the magnetic moment of the ferromagnetic barrier \cite{sara}. In turn, the periodic field of the magnetic component through the interaction with the magnetic moment locks the Josephson oscillations leading to the Buzdin steps in the $IV$-characteristics. The combined effect of both components results in the appearance of  a unique step different from BS and SS. As mentioned above, since it comes from both mechanisms of locking we call this step the composite or Chimera step.

\begin{figure}[tbh]
		\centering
		\includegraphics[width=0.9\linewidth]{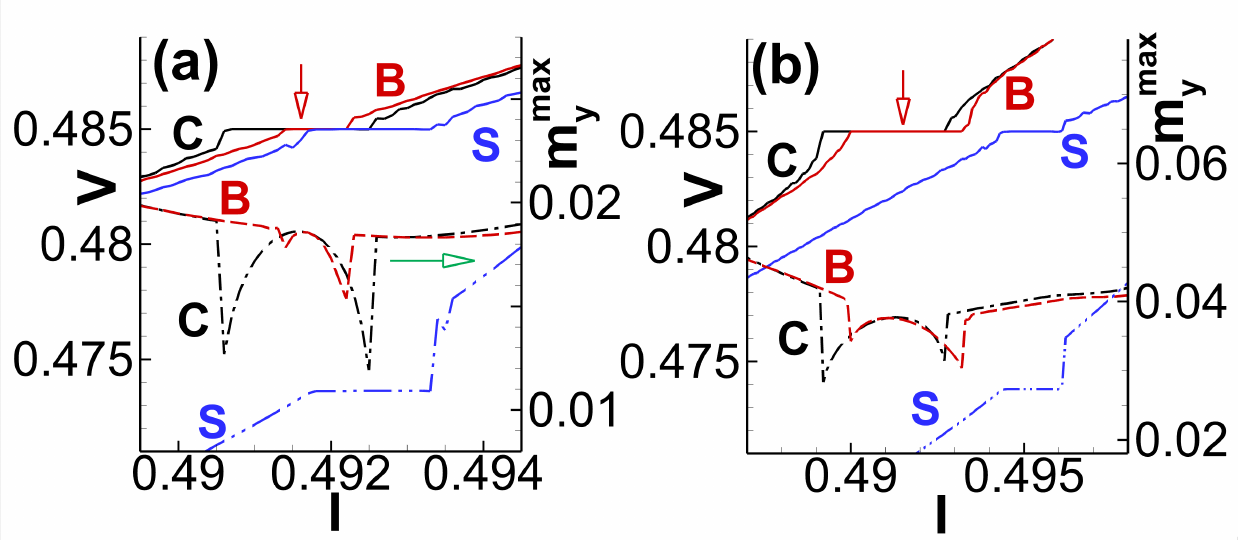}
		\caption{The effects of both radiation components. (a) Parts of IV-characteristics with the Buzdin, Shapiro, and composite steps at $r=0.2$; (b) The same at $r=0.4$.
		}
		\label{Fig4}
\end{figure}

The effect of both the electric and magnetic components of radiation at $r=0.2$, $h_{R}=1$, and  $A=0.01$  is presented in Fig.\ref{Fig4}(a).
It shows the $IV$-characteristics and $m_y^{max}$(I) dependence for three cases: i) in the presence  of two components; ii) the electric component only; and iii) the magnetic component only. We stress that the C-step is not a trivial sum of BS and SS. In particular, we see in its left part that it does not coincide with SS and BS under separate actions. The corresponding $m_y^{max}$(I) dependence shows the bubble structure, which indicates that in this case, we observe also a locking with the changed amplitude of the magnetization precession. It is surprising that the bubble structure continues along the whole CS because, in the second and third cases, we have seen the bubble structure for the BS, and the locking step for SS, where $m_y^{max}$ stays constant along the step. In this figure, the scale of the $m_y^{max}$(I) dependence for the first, and third cases is the same, while in the second case, we have made an arbitrary shift to demonstrate clearly the manifestation of SS.  An important fact here is that the maximal amplitude of the magnetization precession is the same in BS and CS, i.e., CS conserves this characteristic of the magnetic component effect. We stress also that the size of the Shapiro step does not change with the spin-orbit parameter.
	
As we see, BS is smaller than SS at this model parameters. We show in Fig. \ref{Fig2}(c) that the value of BS is growing with $r$. So by variation of the $\varphi_0$ junction and external radiation parameters, we expect the case when BS is larger than SS. This case is shown in Fig.\ref{Fig4}(b) at $r=0.4$. Additionally, SS calculated at $h_{R}=0$ is in the current interval,  which is out of the corresponding interval for the C-step. Like in Fig.\ref{Fig4}(a), we have shifted arbitrarily the $m_y^{max}$(I) dependence for clarity.
	
\begin{figure}[tbh]
		\centering
		\includegraphics[width=0.9\linewidth]{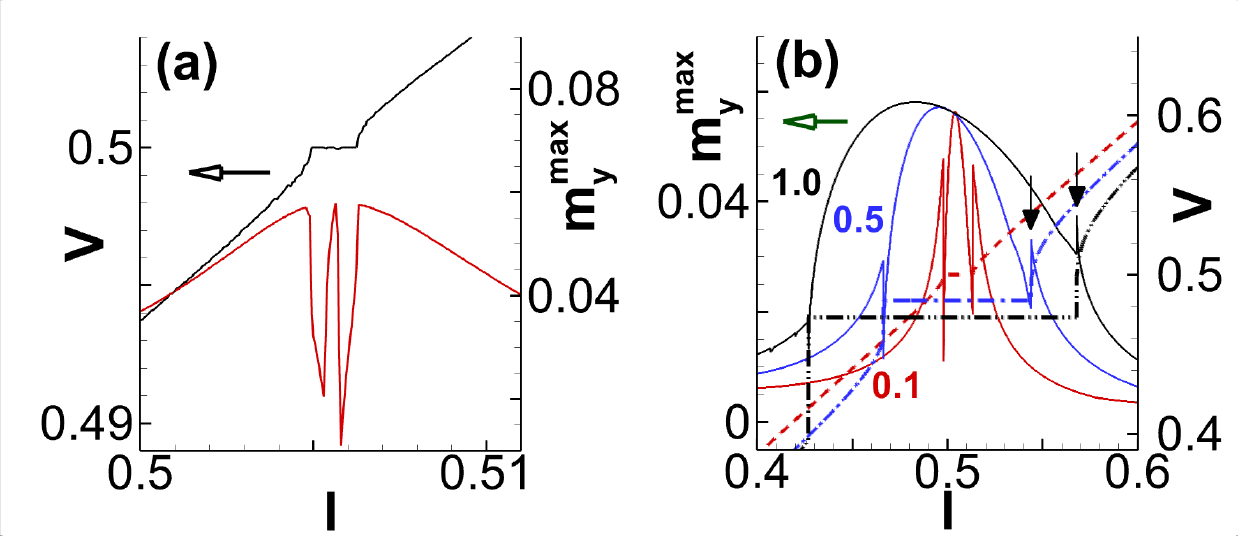}
		\caption{(a) The $IV$-characteristics and  $m_y^{max}$ as a function of $I$ in the case of the magnetic components only for $\omega _R=\omega _F=0.5$. (b) The $IV$-characteristics and $m_y^{max}$ for three cases of different amplitudes of the electric components are indicated by numbers. Other parameters are $G=0.01$, $r=0.2$ and $h_{R}=1$.}
		\label{Fig5}
\end{figure}

So far we have only considered the region near FMR $\omega _F=0.5$ at $\omega _R=0.485$, which leads to the question of how the behavior changes when the frequency of external radiation is equal to the ferromagnetic resonant one. In Fig. \ref{Fig5}(a)  the $IV$-characteristics and $m_y^{max}$  are presented for the magnetic component only ($A=0$ ) at $\omega _R=\omega _F=0.5$. The Buzdin step in the $IV$-characteristics and its locking signature (the bubble structure) in the $m_y^{max}(I)$ dependence can be seen clearly.

In Fig. \ref{Fig5}(b),  the $IV$-characteristics and $m_y^{max}$ are plotted when both radiation components are switched on for three cases of different amplitudes of electric components. We observe a fast increase in the C-step width with the amplitude of the electric component. 
	
Locking is not the only effect that the microwave magnetic field may have on the system. Namely, its direct influence on the magnetic moment of the ferromagnetic layer leads to the Kittel ferromagnetic resonance \cite{kittel-book}. Due to the coupling of the magnetic moment with the Josephson phase, it plays a specific role in the $\varphi_0$ junction. The effective field in the case of the Kittel resonance in normalized units has the following components: $h_{x} = 0$, $h_{y} = G r \sin(\varphi - r m_{y})+  h_{R} \sin(\omega_R t)$, $h_{z} =  m_{z}$.	

\begin{figure}
		\centering
		\includegraphics[width=0.7\linewidth]{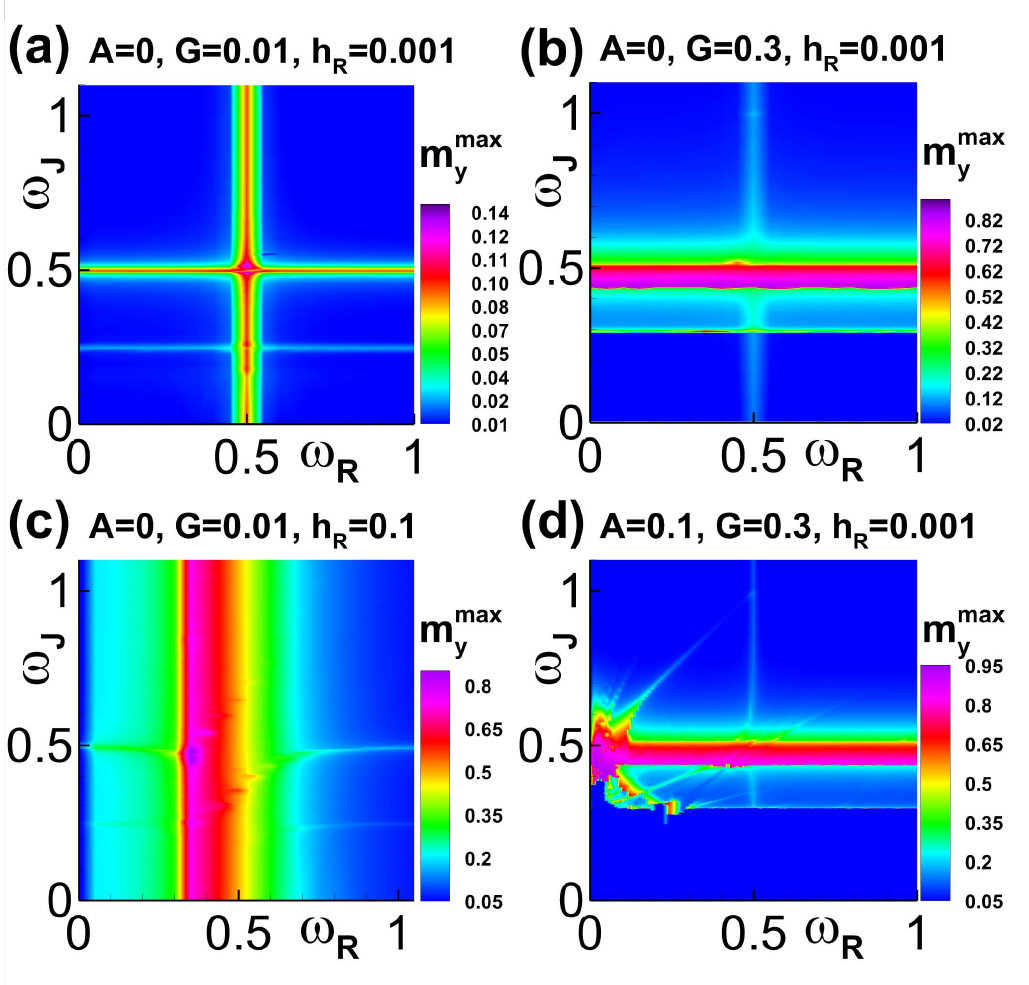}
		\caption{Manifestation of the ferromagnetic Kittel resonance due to the external electromagnetic wave with frequency $0.5$  at $r=0.2$ and different values of A, $h_{R}$ and $G$.}
		\label{Fig6}
\end{figure}

The competition between the Kittel and Josephson FMR at different model parameters is presented in Fig.\ref{Fig6}. 
At rather small values of the magnetic component amplitude $h_{R}$ and Gilbert damping $\alpha$, the Kittel resonance (KR) and Josephson resonance (JR) are centered at $\omega_{R}=\omega_{F}$ and $\omega_{J}=\omega_{F}$, respectively (see Fig.\ref{Fig6} (a)). There is a manifestation of a subharmonic peak at $\omega_{J}=\omega_{F}/2$ pronounced as the horizontal line.  As discussed before, close to the resonance condition $\omega_{J}=\omega_{F}$, we observe a locking of the Josephson oscillation and magnetic moment precession.   With an increase in the coupling constant G, the JR region dominates and the KR region becomes faint (see Fig.\ref{Fig6}(b)). The situation is reversed with increasing $h_{R}$ when the KR dominates.
	
Also, we observe a shift in the resonance frequency related to the effect of magnetic anisotropy. Moreover, the KR region becomes more broadening and nonsymmetric around the resonance frequency. The situation changes dramatically when we take into account both the electric and magnetic components of the microwave field. In this case, a crossed resonance region appears which corresponds to subharmonics and resonances with a combination of $\omega_F$ and $\omega_R$ frequencies. In addition to this, by comparing  Fig.\ref{Fig6}(b) and  Fig.\ref{Fig6}(d), the effect of A is manifested in the JR linewidth. So by changing the frequency and amplitude of the external electromagnetic radiation, one can tune and manipulate the Kittel and Josephson resonances  in the hybrid Josephson junction systems. Experimentally, FMR in the SFS structures has attracted much attention recently ~\cite{Golovchanskiy2023,Golovchanskiy2023,Li2018,Golovchanskiy2020,Eschrig2019}. The  Kittel regime \cite{Kittel1948} can be used to realize a magnetic logic gate through the superconducting phase transition. A gate-controlled time-dependent spin-orbit coupling is proposed and demonstrated by the authors in Ref.\cite{Alidost22}. A dramatic change in current-phase relations and Josephson energy can be seen in this case, even when there is no bias current.  
Using this and the experimental work Ref.\cite{may19}, one can tune the Chimera step by tuning SOC in a hybrid SFS system.
	
One of the ways of testing our results experimentally would be in a superconductor-ferromagnetic insulator-superconductor on a 3D topological insulator which has a strong enough spin-orbit coupling needed for the $\varphi_0$ JJ \cite{bobkova-prb20}. The value of the Rashba-type parameter $r$ can vary in the range $0.1-1$ in such materials as permalloy doped with $Pt$ \cite{hrabec-prb16} or the ferromagnets $MnSi$ and $FeGe$.
Therefore, in the material with weak magnetic anisotropy $K \sim 4\times 10^{-5} KA^{-3}$ \cite{rusanov-prl04}, and a junction with a relatively high critical current density of $(3 \times 10^5 - 5 \times 10^6) A/cm^2$ \cite{robinson-sr12}
the value of the product $Gr$ can easily be in the range $0.01-100$. 
This makes it possible to reach the values used in our numerical calculations for the possible experimental observation of the predicted effects.
	
As a summary, we have found that the magnetic component of radiation brings a series of unique effects in the dynamics of the $\varphi_0$ Josephson junction. 
In particular, unlike the electric component which leads to the Shapiro steps, the magnetic one can interact directly with the magnetic moment of the ferromagnetic layer and due to its coupling with the superconducting phase can lock the Josephson oscillations. This leads to the appearance of the Buzdin steps in the $IV$-characteristics. 
The locking of the magnetization precession is characterized by a bubble structure along the step. The spin-orbit coupling substantially contributes to the amplitude of the Buzdin steps, which exhibits a dramatic increase at frequencies near the ferromagnetic resonance.
If both components drive the system, the presence of two locking mechanisms leads to the Chimera steps.
We stress that BS, SS, and CS have very different properties (amplitude dependence, etc.), which will be studied in detail as a separate work. 
We consider that combinations of the Josephson
and Kittel ferromagnetic resonances and the different types of locking in the $\varphi_0$ Josephson junctions could open up a series of new applications.
	
We thank A. Buzdin, I. Rahmonov, T. Belgibaev for fruitful discussion.
Numerical simulations were funded by  the Russian Science Foundation Project No. 22-71-10022. The work was partially funded by the Ministry of Education, Science and Technological Development of the Republic of Serbia, grant number 451-03-47/2023-01/ 200017 ("Vinča" Institute of Nuclear Sciences, University of Belgrade).
	

\begin{thebibliography}{99}
		
\bibitem {buz08}
A. Buzdin, 
Direct Coupling Between Magnetism and Superconducting Current in the Josephson $\varphi$0 Junction,
\prl \ \textbf{101}, 107005 (2008).
		
\bibitem {kon09}
F. Konschelle and A. Buzdin,
Magnetic Moment Manipulation by a Josephson Current, Phys. Rev. Lett. \textbf{102}, 017001 (2009).
		
\bibitem{shuk22ufn}
Yu. M. Shukrinov,
Anomalous Josephson effect,
Phys. Usp. \textbf{65}, 317 (2022).		
		
\bibitem{szom16}
D. B. Szombati, S. Nadj-Perge, D. Car, S. R. Plissard, E. P. A. M. Bakkers and L. P. Kouwenhoven, Josephson $\varphi_{0}$-junction in nanowire quantum dots,
Nat. Phys., \textbf{12}, 568572 (2016).
		
\bibitem{apr19}
A. Assouline, C. Feuillet-Palma, N. Bergeal, T. Zhang, A. Mottaghizadeh, A. Zimmers, E. Lhuillier, M. Eddrie, P. Atkinson, M. Aprili, H. Aubin, 
Spin-Orbit induced phase-shift in $Bi_{2}Se_{3}$ Josephson junctions,
Nat. Commun. \textbf{10}, 126 (2019).
		
\bibitem{may19}
W. Mayer, M. C. Dartiailh, J. Yuan, K. S. Wickramasinghe, E. Rossi, and J. Shabani, 
Gate controlled anomalous phase shift in $Al/{InAs}$ Josephson junctions,
Nat. Commun., \textbf{11}, 212 (2020).
		
\bibitem{lin15}
J. Linder  and J. Robinson,
Superconducting spintronics,
Nat. Phys.  \textbf{11}, 307 (2015).
		
\bibitem{mel22ufn}
A. S. Mel'nikov, S. V. Mironov, A. V. Samokhvalov, A. I. Buzdin, 
Superconducting spintronics: state of the art and prospect,
Uspekhi Fizicheskikh Nauk, \textbf{192}, 1339 (2022);
Phys. Usp., \textbf{65}, 1248 (2022).
		
\bibitem{bobkova-rev}
I. B. Bobkova, A. M. Bobkov, and M. A. Silaev, 
Magnetoelectric effects in Josephson junctions. 
J. Phys.: Condens. Matter {\bf 34}, 353001 (2022).
		
\bibitem{sara} 
S. A. Abdelmoneim, Yu. M. Shukrinov, K. V. Kulikov, H. ElSamman, and M. Nashaat, 
Locking of magnetization and Josephson oscillations at ferromagnetic resonance in a $\varphi_{0}$ junction under external radiation, 
\prb \ \textbf{106}, 014505 (2022).
		
\bibitem{kittel-book} 
C. Kittel, 
Introduction to solid state physics, (John Wiley and Sons, 2005).
		
\bibitem{Li2018}  L. -L. Li, Y. -L. Zhao, X. -X. Zhang, and Y. Sun, 
Possible evidence for spin-transfer torque induced by spin-triplet supercurrents, 
Chin. Phys. Lett. \textbf{35}, 077401 (2018).
		
\bibitem{Golovchanskiy2023} I.A. Golovchanskiy, N.N. Abramov, O.V. Emelyanova, I.V. Shchetinin, V.V. Ryazanov, A.A. Golubov, and V.S. Stolyarov, 
Magnetization Dynamics in Proximity-Coupled Superconductor-Ferromagnet-Superconductor Multilayers. II. Thickness Dependence of the Superconducting Torque, 
Phys. Rev. Appl.\textbf{19}, 034025 (2023).
		
\bibitem{Golovchanskiy2020} 
I. A. Golovchanskiy, N. N. Abramov, V. S. Stolyarov, V. I. Chichkov, M. Silaev, I. V. Shchetinin, A. A. Golubov, V. V. Ryazanov, A. V. Ustinov, and M. Y. Kupriyanov, 
Magnetization Dynamics in Proximity-Coupled Superconductor-Ferromagnet-Superconductor Multilayers, 
Phys. Rev. Appl. \textbf{14}, 024086 (2020).
		
\bibitem{Eschrig2019} 
K.-R. Jeon, C. Ciccarelli, H. Kurebayashi, L. F. Cohen, X. Montiel, M. Eschrig, T. Wagner, S. Komori, A. Srivastava, J. W. A. Robinson, and M. G. Blamire, 
Effect of Meissner Screening and Trapped Magnetic Flux on Magnetization Dynamics in Thick $Nb/Ni_{80}Fe_{20}/Nb$ Trilayers, Phys. Rev. Appl. \textbf{11}, 014061 (2019)
			
\bibitem{Kittel1948} C. Kittel, 
On the Theory of Ferromagnetic Resonance Absorption, 
Phys. Rev. \textbf{73}, 155 (1948).
		
\bibitem {Alidost22} D. Monroe, M. Alidoust, and I. Žutić, 
Tunable planar Josephson junctions driven by time-dependent spin-orbit coupling, 
Phys. Rev. Applied \textbf{18}, L031001 (2022).
		
\bibitem{bobkova-prb20} 
I. V. Bobkova , A. M. Bobkov, I. R. Rahmonov, A. A. Mazanik , K. Sengupta, and Yu. M. Shukrinov, Magnetization reversal in superconductor/insulating ferromagnet/superconductor Josephson junctions on a three-dimensional topological insulator, 
\prb \ \textbf{102}, 134505 (2020).
		
\bibitem{hrabec-prb16} 
A. Hrabec, F. J. T. Goncalves, C. S. Spencer, E. Arenholz, A. T. N’Diaye, R. L. Stamps, and C. H. Marrows, 
Spin-orbit interaction enhancement in permalloy thin films by Pt doping, 
\prb \ \textbf{93}, 014432 (2016).
		
\bibitem{rusanov-prl04} 
A. Yu. Rusanov, M. Hesselberth, J. Aarts, and A. I. Buzdin, 
Enhancement of the superconducting transition temperature in Nb/permalloy bilayers by controlling the Domain State of the ferromagnet, 
\prl \ \textbf{93}, 057002 (2004).
		
\bibitem{robinson-sr12} 
J. W. A. Robinson, F. Chiodi, M. Egilmez, G. B. Hal\'asz, and M. G. Blamire, 
Supercurrent enhancement in Bloch domain walls, 
Scientific Report \textbf{2}, 699 (2012).	
		
\end{thebibliography}

\end{document}